\documentclass[preprint,12pt]{elsarticle}

\usepackage{amssymb}

\newcommand{\operator}[1]{\bf {#1}}

\newcounter{bla}

\begin{document}

\begin{frontmatter}

\title{Web-Schr{\"o}dinger: Program for the interactive solution of the time dependent and stationary two dimensional (2D) Schr{\"o}dinger equation}

\author[a]{G\'eza I. M\'ark\corref{author}}

\cortext[author] {Corresponding author.\\\textit{E-mail address:} mark@mfa.kfki.hu}
\address[a]{Centre for Energy Research, Institute of Technical Physics and Materials Science, Hungarian Academy of Sciences, P.O. Box 49,
1525 Budapest, Hungary}

\begin{abstract}
Web-Schr{\"o}dinger is an interactive client-server software for the solution of the time dependent and time independent (stationary) Schr{\"o}dinger equation.
The program itself runs on a server computer and can be used through the Internet with a simple Web browser.
Nothing is installed on the user's computer.
The user can load, run, and modify ready-made example files, or prepare her/his own configuration(s), which can be saved on her/his own computer for later use.
Several users can access the program parallelly and each can have independent sessions.
Typical run times are in the second, or minute range.
\end{abstract}


\end{frontmatter}



\section{Introduction\label{sec:intro}}

In non-relativistic quantum mechanics the time dependent Schr{\"o}dinger equation governs the time evolution of the wave function
$\psi(\vec r;t)$,
where $\vec r$ is the position vector
and $t$ is time.
For those states where the probability density
$\rho(\vec r;t) = |\psi(\vec r;t)|^2$
is time independent (stationary), the time evolution is of the form
$\psi(\vec r;t) =
 \varphi(\vec r) e^{i \omega t}
$
where the $\varphi(\vec r)$
{\em stationary wave function} can be calculated from the stationary Schr{\"o}dinger equation,
and $\omega = E / \hbar$,
where $E$ is the energy.
The Schr{\"o}dinger equation is a linear partial differential equation and all the properties of the physical system in question are coded into the
{\operator H} Hamiltion operator.
If the potential is conservative,
then
$\operator H =
 \operator K +
 \operator V
 $,
 where $\operator K$ is the kinetic energy
 operator and $\operator V$ is the
 potential energy operator.
 For local potentials the effect
 of $\operator V$ is
 a simple multiplication
 with the $V(\vec r)$
 potential function.
For the case of localized systems, i.e. where the wave function is localized to a finite spatial region, the $\varphi(\vec r;E)$ stationary wave function is quantized, i.e.
$E \in \{ E_i \}$.
The $E_i$ are the eigenvalues and the $\varphi(\vec r;E_i)$ are the eigenstates of the Hamilton operator.

The wave function is the basic quantity in quantum mechanics, because all measurable quantities – i.e. probabilities of any outcome of any measurement – can be calculated from the wave function.
$\vec r$ is a point in configuration space with dimension $N*D$, where $N$ is the number of particles and $D$ is the degree of freedom
for one particle
$D=$ 1,2, or 3.
The Schr{\"o}dinger equation can be solved analytically only for a few simple examples (e.g. the Hydrogen atom), a numerical solution is necessary for all practically important cases.
Textbook examples – e.g. the rectangular potential well – are mostly given for 1D (one-dimensional) cases, or highly symmetric cases, which can be reduced to 1D (eg. the radial potential).
In practical cases, however, the physical system can rarely be reduced to a 1D problem, moreover some important concepts of quantum mechanics – e.g. the angular momentum – can’t be described or understood in a 1D model.

Hence we created a program, named Web-Sch{\"o}dinger (WS) for the numerical solution of the 2D (two-dimensional) time-dependent and stationary one particle Schr{\"o}dinger equation for local stationary (non time-dependent) localized potentials, where the user can assemble the $V(x,y)$ potential in an interactive graphical editor.
The program then calculates the $\psi(x,y;t)$ time dependent wave function and the $\{ E_i \}$ eigenvalues, together with the $\varphi(x,y;E_i)$ eigenstates.
For the case of the time dependent simulation, the
$\psi(x,y;t_0) = \psi_0(x,y)$
initial wave function has also to be specified.
This can be done also with an interactive editor in WS.
The results are displayed graphically by images and animations.
The user can also determine measurement lines – which correspond in 2D to the measurement planes in 3D – and the program calculates the $I(t)$ probability current and the $T$ transmission probability crossing these measurement lines.

It is an important advantage of the program that it is constructed utilizing a client-server model, where the client (the user interface part) can be accessed through a simple web browser and the server component (the part performing the calculation) runs on a central server computer.
The client-server program is written in such a way, that the system can handle several independent WS sessions paralelly.
These features make the WS software very useful both in studying the behavior of the wave function in models of practical physical systems and in the education of quantum mechanics.

Hartree atomic units are used in all formulas except where explicit
units are given.
SI units are used, however, in all the figures and numerical
data.

A working installation of WS is available here:
 
{\em http://www.nanotechnology.hu/online/web-schroedinger/index.html}.

\section{Calculation methods\label{sec:calc}}

\subsection{Time dependent calculation\label{sec:CalcTimedep}}

Time evolution of the wave function is governed by the time dependent
Schr\"odinger equation:

\begin{equation} \label{eq:TimeDepSchEq}
i \frac { \partial \psi (\vec r;t) } { \partial t }   = 
      {\operator H} \psi (\vec r;t) .
\end{equation}

Given {\operator H} and a $\psi(\vec r;t_0)$ {\em initial state} this equation completely determines $\psi(\vec r;t)$.

\subsubsection{The split-operator FFT method\label{sec:SplitOperatorFFT}}

Formal solution of eq. \ref{eq:TimeDepSchEq} can be written using the time development operator:

\begin{equation} \label{eq:UOperator}
\psi (\vec r;t) = {\operator U} \psi_0 (\vec r;t)
\,\,\,\,\,\,\,\,\,\,\,\,
{\operator U} = e^{ -i {\operator H} ( t - t_0 ) }
\end{equation}

If the potential is conservative, then
${\operator H} = {\operator K} + {\operator V}$  where the
kinetic and potential energy operators do not commute in general,
hence the exponential in eq. \ref{eq:UOperator} can not be factored.
Note, however, that we can decompose~\cite{Fleck1976, Feit1982}
the exponential by the symmetrical unitary product

\begin{equation} \label{eq:SplitOperator}
e ^ { - i ( {\operator K} + {\operator V} ) \delta t }
\approx
e ^ { - i {\operator K} \delta t / 2 }
e ^ { - i {\operator V} \delta t     }
e ^ { - i {\operator K} \delta t / 2 }
\end{equation}

The error of this approximation is
$O \left[ \left( \delta t \right)^3 \right]$.

According to eq. \ref{eq:SplitOperator} the action of the evolution
operator is split into three consecutive steps:
a free propagation for time  $\delta t / 2$,
a potential only propagation for time $\delta t$
and again a free propagation for time  $\delta t / 2$.
If the potential  ${\operator V}$  is local, then its effect is
a simple multiplication with  $V(\vec r)$, hence the effect of the
potential energy propagator
$\exp \left( - i {\operator V} \delta t \right)$  is a multiplication
with  $\exp \left( - i  V(\vec r)  \delta t \right)$.
The effect of the kinetic energy propagator
$\exp \left( - i {\operator K} \delta t / 2 \right)$
is given in  $k$  space by multiplicating the
$\psi (\vec k;t)$  momentum space wave function by
$\exp \left( i {|{\vec k}|}^2 \delta t / 4 \right)$
To utilize this formula it is necessary to calculate the
$\psi (\vec k;t)$  momentum space wave function by fast
Fourier transform (FFT) of  $\psi (\vec r;t)$.
Finally we have to return back to real space by inverse FFT, i.e.:

\begin{equation} \label{eq:KineticPropagation}
e ^ { - i {\operator K} \delta t / 2 }  \psi(\vec r;t) =
{\cal F}^{-1} \left[
	      e ^ { i {|{\vec k}|}^2 \delta t / 4 }
	      {\cal F} \left[ \psi(\vec r;t) \right]
            \right]
\end{equation} 

The evolution of the wave function over a time step  $\delta t$
is calculated in a straightforward way: first
eq. \ref{eq:KineticPropagation} is applied, then the result is
multiplied by  $\exp \left( - i  V(\vec r)  \delta t \right)$,
and finally eq. \ref{eq:KineticPropagation} is applied again.
Convergence towards the exact result is obtained by using a
small  $\delta t$.

The WS software package contains
a direct implementation of the split time FFT method.
The  $\psi(x,y;t)$  wave function is discretized in space and time on a uniform mesh.
The program takes the $V(x,y;t)$ potential and the the initial wave function
$\psi_0(x,y)$ as input then applies a numerical
algorithm representing the time evolution operator
to calculate the  $\psi$  for
the next time increment:

\begin{equation}
\psi_{n+1} (x,y)  =  {\operator U}_{\delta t} \psi_n (x,y)
\end{equation}

where  $\psi_n(x,y) = \psi(x,y;t=n \delta t)$.
The  $\psi_n(x,y)$  wave function is discretized on a
2D mesh,
$
\Psi_{n,(i,j)}    =
   \psi( i \delta x, j \delta z; n \delta t)
$
.

The time advance algorithm for a time step  $\delta t$
is composed of three consecutive operations:

\begin{enumerate}

  \item
  $\Psi_n ^ {\left( 1 \right)} =
    FFT^{-1} \left[
                P_{kinetic}(\delta t / 2) \, FFT\left( \Psi_n \right) 
             \right]
  $

  where
  $P_{kinetic}(\delta t / 2) =
    \exp \left( i {|{\vec k}|}^2 \delta t / 4 \right)$
  is the kinetic energy propagator for time  $\delta t/2$

  \item
  $\Psi_n ^ {\left( 2 \right)} =
    P_{potential} (\delta t) \Psi_n ^ {\left( 1 \right)}$

  where
  $P_{potential} (\delta t) = \exp \left( - i  V(\vec r)  \delta t \right)$
  is the potential energy propagator for time  $\delta t$

  \item
  $\Psi_{n+1} =
    FFT^{-1} \left[
                P_{kinetic}(\delta t / 2) \,
		FFT\left( \Psi_n ^ {\left( 2 \right)} \right)
             \right]$

\end{enumerate}

If we are not interested in the value of  $\Psi_n$  for every time step
then we can combine the 3. operation of step $n$ with the 1. operation
of step $n+1$ into a single operation of a kinetic energy propagator
for a full time step  $\delta t$.
Since most of the computing time is spent in calculating the
FFT-s, this trick decreases the total computing time to
nearly half.

$\delta x,y$  has to be fine enough to sample the smallest
de Broglie wave length found in the WP and also
to sample the potential  $V(\vec r)$  with enough detail.
For example, for a  $E_k = 5$~eV  kinetic energy  $\lambda_{de\ Broglie} = 0.55$~nm.
$\delta x,y$  and  $\delta t$  can not be chosen independently
because the kinetic energy propagator contains a pure imaginary
quantity in the exponent and this exponent has to be $< 2 \pi$
to prevent unphysical aliasing effects.
The largest reciprocal lattice vector  $k_{x,y}^{max} = 2 \pi / \delta x,y$,
hence the condition for  $\delta t$  is:

\begin{equation}
\delta t < \frac{4}{\pi} \frac{(\delta x,y)^2}{D}
\end{equation}

where $D$ is the number of dimensions, which is  $D=2$    in WS.

Most of the computing time is spent by calculating the
Fourier transforms.
CPU time of the Fast Fourier Transform (FFT) algorithm is
scaling with  $n \log n$, hence the split operator FFT method
can handle large problems.
The FFT algorithm can be parallelized very efficiently.
The ideal computer architecture to calculate FFTs is the
shared memory architecture.

\subsubsection{Construction of the initial state -- the wave packet dynamical method
	    \label{sec:InitialState}
	   }
	   
\begin{figure}
	\centering
		\includegraphics[width=12 cm]{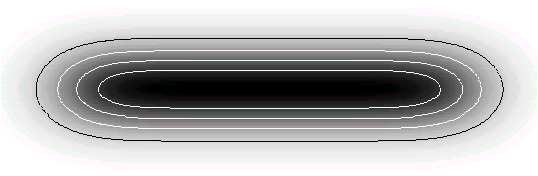}
	\caption{\label{fig:tpw2d}
	Probability density of a 2D truncated plain wave wave packet.
	White shows the zero density and black the maximal density.
	The isodensity lines show that density of this wave packet is indeed constant along a finite horizontal region.
    }
\end{figure}

Erwin Schr\"odinger introduced the concept of
wave packets~\cite{Schrodinger1926WP} (WPs) to bridge the gap between
classical and quantum descriptions of nature.
The wave packet dynamical (WPD) method~\cite{Garraway1995WPD,Varga2002}
is a scattering experiment inside the computer: an incoming WP
is "shot" into the localized potential under investigation
and the time development of the WP is calculated by solving
the time dependent Schr\"odinger equation
(or with some other method).
The "results" of the WPD calculation are: i) the
time development of the WP itself -- it gives insight
into the details of the time dependent dynamical process, e.g. into resonances; ii) the "outgoing WP" (or scattered WP), i.e. the WP in the far space (far away from the localized potential); and iii) selected measurables calculated from the
scattered WP.
These measurables are those quantities that can be measured in a real
scattering experiment (at least in principle).
{\em Fig.~\ref{fig:tpw2d}} shows the schematics of a WPD calculation.

\begin{figure}
	\centering
		\includegraphics[width=12 cm]{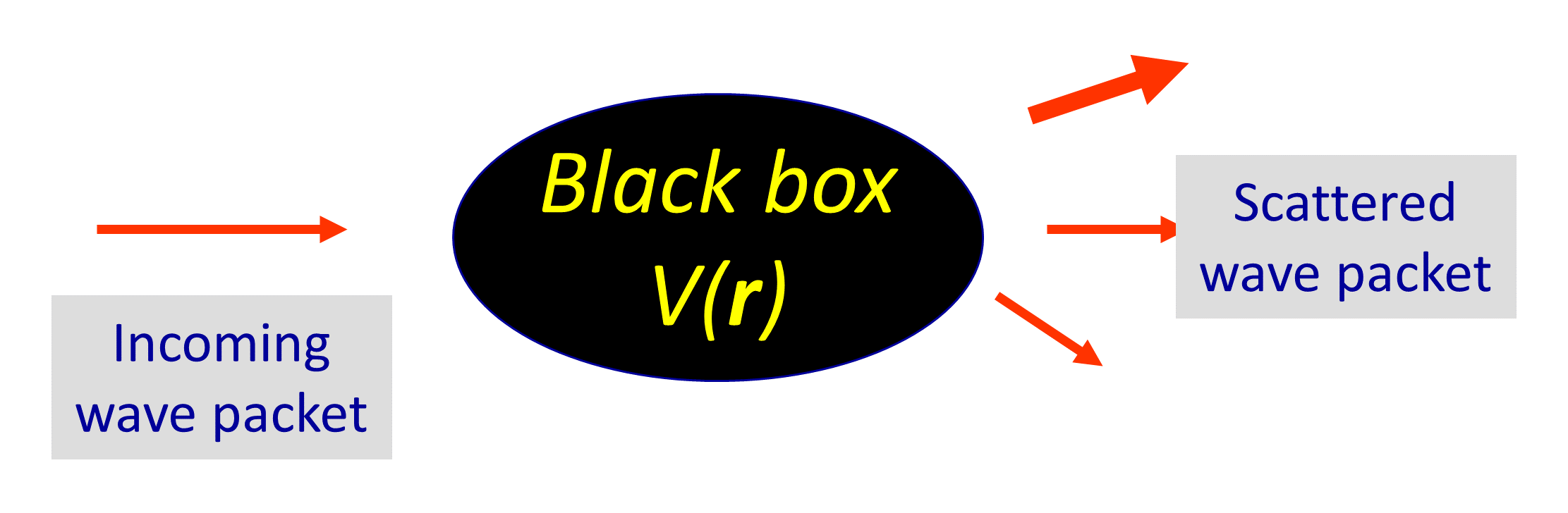}
	\caption{\label{fig:WPD}
Principe of the wave the wave packet dynamical method.
It is a scattering experiment inside the computer.
We have a localized physical system
(a "black box"), characterized by a local $V(\vec r)$ potential.
The incoming wave packet is scattered on this potential.
The red arrows on the right hand side represent the scattered wave packet.
We drew several arrows with different thicknesses, symbolising that the incoming wave packet is scattered into different directions with different amplitudes (and phases).
    }
\end{figure}

In the recent years we applied the WPD method in the analysis of electronic transport in
various nanosystems~\cite{Mark2016NATO}, e.g.
the imaging process of the STM\cite{Binnig1982STM} (Scanning Tunneling Microscope)
 of carbon nanotubes~\cite{Mark1998PRB,Mark2000PRB,Mark2004PRB}, nanotube networks~\cite{Kvashnin2015Bilayer}, and graphene~\cite{Mark2012PRB}, as well as for
graphene grain boundaries~\cite{Vancso2014DisorderGB}.
Later we investigated the WP propagation on other 2D materials, e.g. single layer MoS2~\cite{Mark2017MoS2} sheets.
WPD calculations are also actively used for topological insulators~\cite{Michala2020wavepacket}, this work is under progress.
The above mentioned calculations were all done with the {\tt sfnd} Fortran program which is part of this package.
In the case of an STM
we have a tunnel junction between
two electrodes, one of the electrodes is a sharp movable tip, another is a flat sample.
In the WPD simulation of the STM tunnel junction, the WP is approaching the
STM tunnel junction from inside of the tip bulk and we study the tunneling of this WP
into the other electrode, the sample.
Given the initial state  $\psi_0 (\vec r;t)$  and the localized
potential  $V(\vec r)$,  the  $\psi (\vec r;t)$  wave
function is calculated from the time dependent Schr\"odinger
equation.

WPD is a conceptionally simple and easy to implement method, it contains
no perturbative approximation but includes all
interference~\cite{Sautet1993Interference} and multiple scattering
effects and thus it is capable of providing results comparable
with advanced tunneling theories~\cite{Cerda1997Green}
when applying a properly chosen model potential.
Inclusion of
multiple scattering and interference effects is important
for modeling the resonant tunneling process~\cite{Mark2004PRB} arising
because of the existence of two tunnel gaps~\cite{Mark1998PRB},
as is often the situation in STM experiments.
When, for example, the sample is a carbon nanotube on a graphite support~\cite{Mark2004PRB}, we have two tunnel gaps: one between the STM tip and the nanotube, another between the nanotube and the graphite support surface.

In free space the time evolution of a WP is quite simple:
it is a translational movement with constant  $\vec v_0$  velocity
and a spreading~\cite{Mark1999Spreading}.
For a complicated potential, however, WPD reveals
a whole lot of interesting phenomena, e.g. the WP
is split into several parts in space and time, periodic or
quasi periodic motion occurs
(so called quantum revivals~\cite{Kaplan2000Revival}, see the {\tt quantum\_revival} example in WS), etc.

A WP represents a quantum system that is localized
in its position coordinate.
From the Schr\"odinger equation we obtain equations of motion
for the  $\left< {\vec r} \right>$,  $\left< {\vec k} \right>$
expectation values of the position and momentum.

\begin{eqnarray} \label{eq:WPEquationR}
\frac{\partial \left< {\vec r} \right>}{t}	&=&
\left< {\vec k} \right>
\\
\label{eq:WPEquationK}
\frac{\partial \left< {\vec k} \right>}{t}	&=&
- \left< \nabla V(\vec r) \right>
\end{eqnarray}

Solutions of these equations give the trajectory of the WP.
Note that in general

\begin{equation} \label{eq:QCTrajectoryUnequality}
\left< \nabla V(\vec r) \right> \not=
\nabla_{\left< \vec r \right>} V(\vec r)
\end{equation}

i.e. the quantum mechanical trajectory may differ from the classical one
for which equality holds in eq. \ref{eq:QCTrajectoryUnequality}.

The width of the WP in coordinate and momentum space is:

\begin{eqnarray} \label{eq:DeltaX}
   \Delta r_i				& = &
      \sqrt { \left< {r_i}^2 \right> -
	      { \left< {r_i} \right> } ^ 2
	    }
   \\		 
   \Delta k_i	\label{eq:DeltaK}	& = &
      \sqrt { \left< {k_i}^2 \right> -
	      { \left< {k_i} \right> } ^ 2
	    }
\end{eqnarray}

where  $i \epsilon \{ x, y, z \}$.
According to the Heisenberg inequality
$\Delta r_i  \cdot \Delta k_i  \ge  1 / 2$  for all WPs.
The equality holds for the Gaussian WP defined as:

\begin{equation} \label{eq:Gaussian}
   \psi_{Gauss} ( \vec r ; a, \vec r_0, \vec k_0 )	=
      \left( \frac {2} { \pi a^2 } \right) ^ \frac{D}{4}
      \cdot
      \exp{ \left( i \vec k_0 \cdot \vec r \right) }
      \cdot
      \exp{ \left( - \frac { \left| \vec r - \vec r_0 \right|^2 } { a^2 } \right) },
\end{equation}

where $\vec r_0$ is the initial position, $\vec k_0$ is the initial momentum, and
$D$ is the dimension, which is $D=2$ in WS.
The width of the Gaussian is  $\Delta r_i = a / 2$ .
In the general case the width can be different in each direction,
i.e. different  $a_x, a_y, a_z$  values can be used
in eq. \ref{eq:Gaussian}.
The momentum space wave function,
$\psi(\vec k) = {\cal F} [ \psi(\vec r) ] (\vec k)$ 
is also a Gaussian with  $\Delta k_i = 1 / a$ .

If the envelope of the WP is not a Gaussian, but some other function, e.g. a square, or a triangle, then for a given $\Delta x$ width the $\Delta k$ momentum space width will be considerable larger than for the Gaussian.
Hence such WPs do spread much faster~\cite{Mark1997Shape}, than a Gaussian WP.

In free space  (${\operator V} = 0$)  the momentum distribution does not change
(see eq. \ref{eq:WPEquationK}), the momentum wave function is
only multiplied by a phase factor:

\begin{equation}
\psi(\vec k;t) =
  \psi_0(\vec k)
  \exp{\left( -i \frac{|\vec k|^2}{2} t \right)} \, .
\end{equation}

Hence the coordinate wave function is the convolution of the
initial wave function with the free time eopagator:

\begin{eqnarray}  \label{eq:EvolutionWithFreePropagator}
   \psi(\vec r;t)		&=&
      P_{Free}(\vec r;t) * \psi_0(\vec r)
   \\
   P_{Free}(\vec r;t)		&=&
      \frac {1} { \sqrt{ 2 \pi t } }
      \exp{ \left( -i \frac{\pi}{4} \right) }
      \exp{  \left( i \frac{|\vec r|^2}{2 t} \right) }
\end{eqnarray}

In the WPD method
(see Section \ref{sec:InitialState})
the time dependent Schr\"odinger equation is solved on
a finite spatial region.
This finite region includes the localized part of the potential,
the potential is zero (or constant, or at least translation symmetric)
outside this region.
We have to solve the time dependent Schr\"odinger equation for
the finite region in such a way that the solution should be the same
as if we solved the time dependent Schr\"odinger equation
for the whole space and then we cutted the part corresponding
to the finite region from the total wave function.
In other words the finite region is a window
to the whole solution.

To eliminate the effect of the particular WP shape on the
resulting tunneling probability the width of the WP has to be
larger than the
size of the localized
scattering potential.
This would require the use of a Gaussian WP with fairly large
$\Delta x$  which would subsequently require a fairly large spatial mesh.
To avoid this difficulty we constructed a WP called
{\em truncated plane wave}~\cite{Mark2000PRB} which has a plateau of
constant probability density larger than the scattering region.
{\em Fig.~\ref{fig:tpw2d}}  shows a grayscale plot
of a truncated plane wave WP.
Such a WP can be constructed as a convolution of a Gaussian with a
square window function.
To compensate for the effect of the distortion of the plateau
during the time development of the WP a backward time propagator
is used to construct the initial state:

\begin{eqnarray} \label{eq:Psi0}
\psi_0 (x,y) &=& \psi_0^x(x) \cdot \psi_0^y(y)
\\
\psi_0^x (x) &=&
   A_x
   \cdot
   \hat P_{t_{x_0}}
   \left[
          \int_{d_{x1}}^{d_{x2}}
	     \exp{ \left( - { { ( x' - x )^2 } \over { a_x^2 } } +
			    i k_x x
		   \right)
		 }
	  \, dx'
   \right]
    \\
\psi_0^y (y) &=&
   A_y
   \cdot
   \hat P_{t_{y_0}}
   \left[
          \int_{d_{y1}}^{d_{y2}}
	     \exp{ \left( - { { ( y' - y )^2 } \over { a_y^2 } } +
			    i k_y y
		   \right)
		 }
	  \, dy'
   \right]
\end{eqnarray}

where $d_{x1}$, $d_{x2}$, $d_{y1}$, $d_{y2}$,
and
$A_x$, $A_y$ are normalization constants.
By the free space propagators $\hat P_{t_{x_0}}$,
$\hat P_{t_{y_0}}$ the
truncated plane wave is backward propagated in time by an amount
of  $t_{x_0} = ( y_0 - y_{interf} ) / v_y$ and
$t_{y_0} = ( x_0 - x_{interf} ) / v_x$
where $ ( x_{interf}, y_{interf} ) $
is the position of the edge of the scattering region,
and  $(v_x,v_y)=(k_x,k_y)$  is the group velocity.

\subsubsection{Choice of the measurables\label{sec:Measurables}}

The method of analyzing the resulting large
wave function dataset basically relies on calculation of integrals
of certain quantum mechanical observables derived from the wave function
on carefully chosen subspaces.
As a first step two important observables are calculated from
the wave function:
the  $\varrho(x,y;t) = |\psi (x,y;t)|^2$
probability density and the  $\vec j(x,y;t)$
probability current density.

$j(\eta;t)$, the perpendicular component of the
$\vec j(x,y;t)$  probability current
density flowing across selected  $\cal L$ {\em measurement lines},
gives the probability current density crossing those lines as the
function of time, where  $\eta$  is the parametric coordinate
(inner coordinate) of the line, which is simply
the arc length along the line.
$\int j(\eta;t) d \eta$  gives the  $I(t)$  probability
current crossing the particular measurement line as the function of time.
By calculating the indefinite integral  $T(t) = \int_0^t I(t') dt'$, 
we determine the transmission probability vs. time, i.e. the portion of the WP
that has crossed the measurement line until time  $t$.
The  $T(t=\infty)$  asymptotic value gives the total transmission probability
for that line.

Surface integral of  $\varrho(x,y;t)$  for selected
$\cal A$ {\em measurement areas} gives the probability of finding
the particle in those areas at the instant $t$,
$P(t) = \int \varrho(x,y;t) dA$.
Integration for the whole space gives  $P(t) \equiv 1$.
Integrating  $P(t)$  in time~\cite{Iannaccone1995DwellTime} gives the
{\em dwell time}~\cite{Buttiker1983LarmorClock,Leavens1989DwellTime},
the average time spent by the particle in $\cal A$:

\begin{equation} \label{eq:DwellTime}
t_{spent} =
  \int_0^{\infty} \left(
			  \int_{\cal A} \varrho(x,y;t) dA
		   \right)
  dt \, .
\end{equation}

By exchanging the integrals in eq. \ref{eq:DwellTime} the dwell
time can be written as a surface integral:

\begin{equation} \label{eq:DwellTimeDensity}
t_{spent} =
  \int_{\cal A}  \tau(x,y) dA
\end{equation}

where

\begin{equation}
\tau(x,y) =
  \int_0^{\infty} \varrho(x,y;t) dt
\end{equation}

is the {\em dwell time density}.

\subsection{Time independent calculation\label{sec:CalcStationary}}

The time independent Schr{\"o}dinger equation

\begin{equation}
    {\operator H} \varphi(\vec r;E) = E \varphi(\vec r;E)
\end{equation}

is an eigen equation, and for localized wave functions it has solutions only for discrete $E_i$ values.
For the WS we would like to find a method for the efficient calculation of the low-lying (but can be a few hundreds) bound-states in two dimensions.
The potential $V(x,y)$  is given on an equidistant spatial grid.
Direct diagonalization methods, with computational effort proportional to $N^3$ ($N$ is the total number of grid points), are impractical in this case.
We were able to adapt the AEDR program\cite{Janacek2008AEDR,Chin2009AEDR}, published in CPC Program Library for our needes.
This software utilizes also a split operator method, similar to that presented in Section \ref{sec:SplitOperatorFFT}, but in imaginary time (the so called {\em diffusion algorithm}). Very briefly, the method works as follows:

\begin{enumerate}

    \item 
    Start with a set of trial wave functions
    $\{ \chi_i^{(l=0)}(\vec r)  \}_{i=1}^n$.
    
    \item
    Apply the imaginary time evolution operator
    ${\hat \tau}(\epsilon) = \exp{ \left( -\epsilon {\operator H} \right)}$
    on the set of trial states,
    $\{ \eta_i^{(k+1)} (\vec r) = {\hat \tau} (\epsilon) \chi_i^{(k)} (\vec r) \}_{i=1}^n$.
    
    \item
    Orthonormalize the the states
    $\{ \eta_i^{(k+1)} (\vec r) \}_{i=1}^n$,
    which produces a new set of trial states
    $\{ \chi_i^{(k+1)} (\vec r) \}_{i=1}^n$.
    
    \item
    Repeat steps (2) and (3) until the set $\{ \chi_i(\vec r)  \}_{i=1}^n$ converges, thus giving the
    $\{ \varphi_i(\vec r)  \}_{i=1}^n$ eigenstates and $E_i$ eigenvalues.
    
\end{enumerate}

All the details of the method are given in~\cite{Janacek2008AEDR,Chin2009AEDR}, here we just briefly note that developing any superposed state in imaginary time converges to the lowest energy component of the particular superposed state.
The orthogonalization step ensures that the algorithm eventually converges to the 
different eigenstates.
Without this step all of the trial states would converge to the ground state.

\subsection{Boundary condition\label{sec:boundary}}

As was shown in Sec. \ref{sec:CalcTimedep} and Sec. \ref{sec:CalcStationary}, both the time dependent- and the time independent calculations rely on Fourier transform methods.
This introduces a periodic boundary condition into the calculation, as if the whole 2D plane were tiled with the calculation window.
If we want to avoid interference effects between the adjacent tiles, it is necessary to keep our wave function far enough from the boundaries.
This can be accomplished by a careful choice of the parameters.

\section{The data flow and the menu system\label{sec:dataflow}}

\begin{figure}
	\centering
		\includegraphics[width=12 cm]{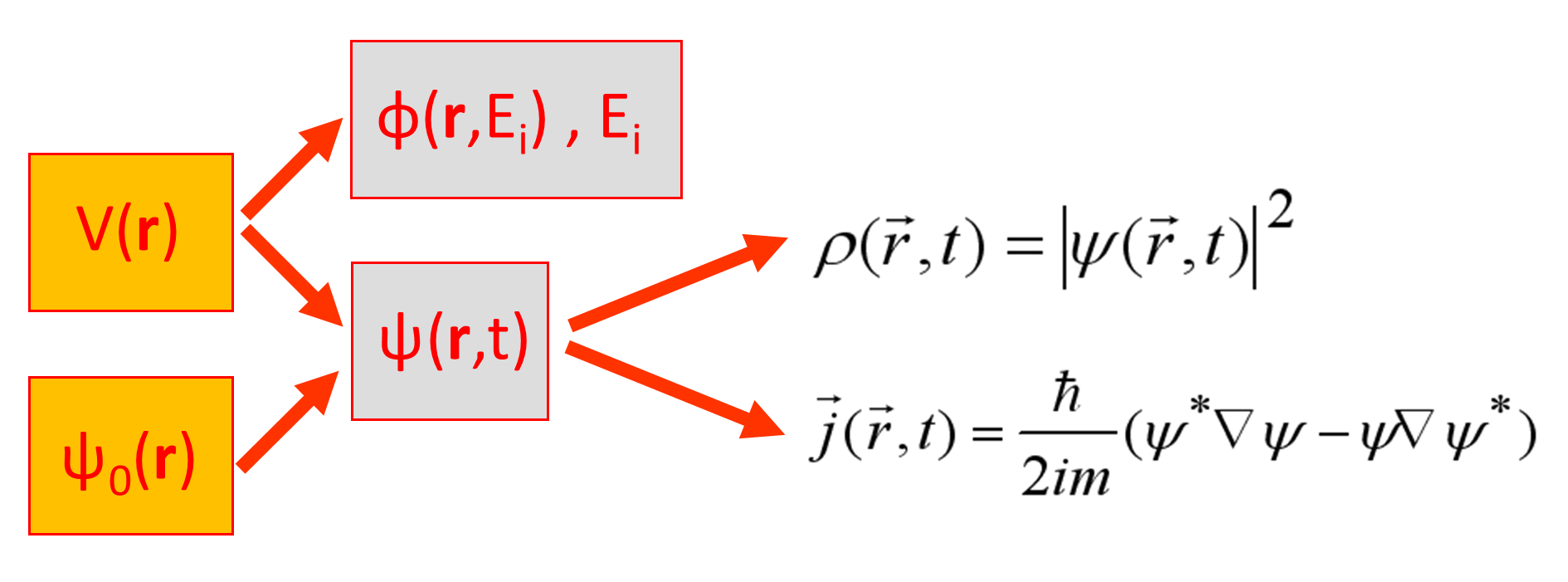}
	\caption{\label{fig:calcflow}
	Calculation model in Web-Schr{\"o}dinger.
    }
\end{figure}

{\em Fig.~\ref{fig:calcflow}} shows the calculation model in Web-Schr{\"o}dinger.
The main input functions are the $V(x,y)$ potential and the $\psi_0(x,y)$ initial wave function.
Both are necessary for the time dependent calculation, but only the potential is necessary for the time independent calculation.
The program then calculates the time dependent wave function or the eigenstates and eigenvalues, and the requested measurables.

Apart from the above mentioned two input functions, the user has to specify some important parameters, such as the spatial and temporal grid, the total time to be calculated, the requested number of eigenstates, the places of the detector lines, etc.
There are default values for all the input parameters.

\begin{figure}
	\centering
		\includegraphics[width=12 cm]  {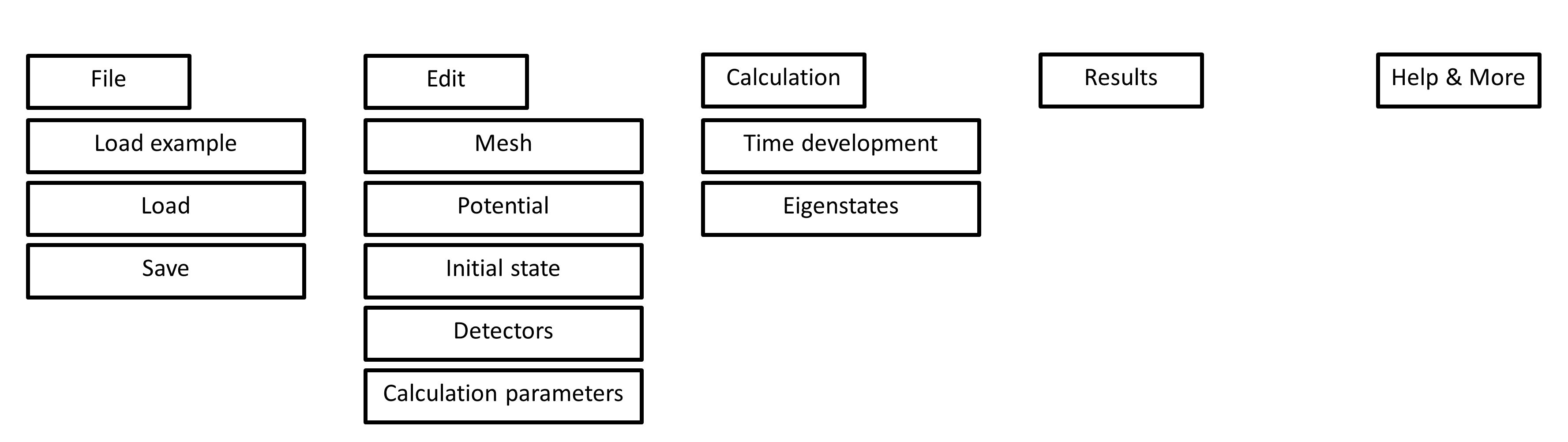},
	\caption{\label{fig:menu}
	Menu system of Web-Schr{\"o}dinger.
    }
\end{figure}

The above explained data flow is mapped into the menu system, cf. {\em Fig.~\ref{fig:menu}}, in a straightforward way.

\subsection{The project file\label{sec:PRJFile}}

All input data necessary for the calculation (cf. Sec.~\ref{sec:dataflow}) are stored in a project file.
The project file is a simple object oriented text file, which contains the parameters in human-readable form.
The user can use the default parameters, load the example project files provided with the program, modify the parameters through the menu system, or load a self constructed project file.
It is also possible to modify or construct project files with an external editor, or with an external program, which makes it possible to perform complicated calculations, and/or utilize potentials created by other software.

\subsection{The menu system\label{sec:menu}}

{\em Fig.~\ref{fig:menu}} shows the two level menu system.

\begin{itemize}

    \item
    "File" menu.
    The File menu makes it possible to read and write the project files.
    
    \begin{itemize}
    
        \item
        "Load example".
        Several examples stored on the server for both the time dependent and the time independent calculation, e.g. tunneling, band structure, bound states in a box, etc.
        See Sec. \ref{sec:examples} for details.
        After loading the chosen example file, the user can go straight to perform a calculation by choosing items from the "Calculation" menu, or edit the example file by the "Edit" menu.
        
        \item
        "Save".
        This creates a project file, i.e. saves the current values of all of the parameters into a text file on the user's computer.
        (S)he can modify it with a text editor, or with an external software.
        
        \item
        "Load".
        This loads a project file from the user's computer into WS. If the file is not properly formatted, the program displays an error message.

    \end{itemize}
    
    \item
    "Edit" menu.
    The Edit menu makes it possible to change all the parameters.
    
    \begin{itemize}
        \item
        "Mesh".
        Here we can define the $x$, $y$ spatial mesh, by giving $N_x$ and $N_y$, the number of mesh points, as well as $s_x$ and $s_y$,
        the size of the calculation window (in Electronvolt).
        Also we can change $V_0$ the value of the background potential here.
        Note, that as a general rule of thumb, a smaller mesh is enough for a good accuracy in the time independent calculation, because of the higher degree split-operators used there.
        See the example files for examples.
        
        \item
        "Potential".
        Here we can define the $V(x,y)$ potential.
        The potential is composed of any number of 2D objects, which can be of three kinds: {\em circle}, {\em plane}, and {\em rectangle}.
        All of these objects has two types of parameters: {\em geometrical parameters} and {\em functional parameters}.
        The geometrical parameters determine the location and size of the object, the functional parameters define, how the potential value changes within the object.
        The present version of WS uses linear and quadratic potential functions within the objects.
        By calculating a step-wise, a linear, or a quadratic interpolation with an external program, creating an appropriate project file, and loading it into WS, any desired $V(x,y)$ potentials can be utilized.
        (The higher the degree of interpolation, the less objects are necessary for a sufficiently accurate interpolation.)
        Simple potentials, however -- like the ones provided in the example file directory (tunneling, hardcore, step, gravity, etc) -- can be easily assembled manually with the graphical editor in this menu point.
        
        \item
        "Initial state".
        Here we can define the initial state $\psi_0(x,y)$ for the time dependent calculation.
        (The $\psi_0$ function is not used in the time independent calculation.)
        The initial state is a {\em truncated plane wave} wave packet (cf. Section \ref{sec:InitialState}), which is a convolution of a Gaussian with a 2D rectangle.
        
        \item
        "Detectors".
        Here we can define vertical and horizontal detector lines.
        The program calculates for each detector the perpendicular component of the $\vec j$ probability current crossing the detector line, and from this the $I(t)$ probability current and the $T(t)$ transmission probability as the function of time.
        
        \item
        "Calculation parameters".
        The small table (an inpot form) displayed in this menu item is divided into two parts: (i) parameters for the time dependent calculation: the calculation time step $\Delta t$, the total number of timesteps requested $N_t$, and the display time step $N_{disp}$ and (ii) parameters for the time independent calculation, which include only $N_{eig}$, the number of eigenstates requested.
        
    \end{itemize}
    
    \item
    "Calculation" menu.
    This starts the actual calculation on the server. The calculation begins when the user hits the "RUN" key.
    After starting the calculation, the program displays a link.
    The user can click on the link immediatelly in order to see the results, or save the link and enter it later into the address line of the web browser.
    This technique is useful when the user launches a longer calculation.
    The CPU time of a calculation is presently not limited, but the number of the spatial- and temporal mesh points are.
    This means that a simple calculation is completed within several seconds, but complicated calculations can be as long as 10, or 20 minutes.
    
    \begin{itemize}
    
        \item
        "Time development".
        This launches the time development calculation (cf. Sec.~\ref{sec:CalcTimedep}).
        The sequential nature of the time development calculation (one time step after another) makes it possible to display the results sequentially, while the calculation is still running.
        The program does just that, it displays thumbnail image snapshots of the $\varrho(x,y;t)$ probability density function, together with the $V(x,y)$ potential, as composite images.
        Yellow is the probability density, blue is the potential.
        Each snapshots are normalized individually.

        \item
        "Eigenstates".
        This launches the time independent calculation, i.e. the solution of the stationary Schr{\"o}dinger equation (cf. Sec.~\ref{sec:CalcStationary}).
        When the calculation is finished, the program displays the $E_i$ eigenvalues and the $\varphi(x,y;E_i)$ eigenfunctions.
        The eigenfunctions are shown as color coded composite RGB images, together with the $V(x,y)$ potential: green means a negative function value, red a positive function value, and blue is the potential value.
        Each snapshots are normalized individually.
        
    \end{itemize}
    
    \item
    "Results" menu.
    This menu item provides a more detailed analysis of the calculation results for the case of the time dependent calculation.
    It displays similar composite images of the probability density and the potential as shown in menu item "Calculation/Time development", with the following differences: (i) the probability density is globally normalized for the total time scale, (ii) we use a nonlinear gray scale here, which is better suited for highliting small probability densities, and (iii) the detector lines (if any) are superimposed to the composite images as red line segments.
    After the web browser displays all of the images of the time series, the server creates a movie from these images and the user is presented a link to this GIF animation.
    If the user specified any detectors, the $I(t)$ probability currents are also displayed, as $I-t$ plots, together with the $T$ transmission probability values for each detectors.
    
    \item
    "Help \& more".
    This shows a brief hypertext user manual of WS, together with a description of all of the examples provided, and a list of external links and references.
    
\end{itemize}

\section{Structure of the program\label{sec:ProgramStructure}}

The software contains five layers and these are written in different languages.

\begin{itemize}

  \item
  The solution of the time dependent and the stationary 2D Schrödinger equation, as well as the calculation of the measurables, are performed by Fortran programs on the server
  ({\tt sfnd} and {\tt 2dsch}).
  
  \item
  An interface layer, written in PERL and running also on the server, connects the calculation programs to the user interface. This PERL program constucts the different graphical and textual screens of WS and codes them into HTML language. It also constucts input forms in HTML, then reads and interpets the user input.
  
  \item
  A standard web server software – e.g. Apache – running on the server computer manages the communication between the client- and server sides of WS.
  
  \item
  The user interface of WS is presented to the user by a standard web browser program (Chrome, Internet Explorer, etc).
  
  \item
  Basic checking of the user input is performed on the client slide, utilizing Javascript scripts, prior to sending the input to the server.

\end{itemize}

The basic data flow in WS can be summarised as follows:

\begin{enumerate}[1.]

  \item
  The PERL program running on the server constructs a screen in HTML, passes it to the web server program, and then the screen is sent to the client computer and presented to the user on the client computer by the web browser.
  
  \item
  This screen contains clickable elements and, in some cases, HTML forms containing user input elements (choice lists, text boxes, etc).
  The user provides input in the forms then clicks on some of the links on the page,
  eg. a SUBMIT button, or some menu items.
  
  \item
  Upon pressing the link, the web bowser program (running on the client computer) send the data to the web server program (running on the server computer), which transfers the user input to the PERL program, and that evaluates the input, in some cases calls the calculation Fortran program, then constructs a new screen in HTML.
  
  \item
  GOTO 1.

\end{enumerate}

WS is implemented in such a way, that it can handle several parallel sessions, e.g. in a student laboratory environment.
This is accomplished by assigning a unique session id number to each of the sessions.
When the user first invokes WS, the PERL scipt generates a new id number and this id number is sent together with each user input screen to the web browser.
When the user finishes with the particular input screen, completes the input and finally hits one of the links on the screen, the id number is sent back to the server, together with the contents of the form.
With the help of the id number the server can identify the particular session and reacts accordingly.
All intermediate data and status information of each of the sessions are stored on the server using the session id as a key.

\section{Installation\label{sec:inst}}

The typical installation of WS contains one server computer and several client computers.
Nothing particular has to be installed on the client computers, only a web browser with JavaScript enabled and reliable Internet connection are necessary, which is a default configuration for most of today's personal computers and even on smartphones.

In a lucky case the installation of WS can be as simple as unpacking the installation GZIP file, copying the contents of the directories to appropriate places on the server, and adjusting the URLs in some of the files
(see Sec. \ref{sec:InstProc}).
In more typical cases it may also involve checking of the system requirements (cf. Sec. \ref{sec:SystemRequirements}), installing some libraries and auxiliary software (cf. Sec. \ref{sec:required}), and compiling the Fortran programs (cf. Sec. \ref{sec:programs}).
See Sec. \ref{sec:InstProc} for details.

\subsection{System requirements\label{sec:SystemRequirements}}

The server side software of WS were developed and tested on a Linux computer.
It was not tested on a Windows server, but it should work under the CYGWIN environment~\cite{CYGWIN} without much alterations.

The server computer must meet the following requirements:

\begin{enumerate}

    \item
    Web server software
    
    \item
    Running of CGI scripts enabled on the Web server.
    CGI (Common Gateway Interface~\cite{CGIWikipedia}) is an interface specification for web servers to execute programs
    
    \item
    Stable internet connection between the server computer and the client computers.
    
    \item
    If the particular installation of WS is intended for general use, e.g. when it is open for any connection from the global Internet, it is important to keep an eye on security.
    Keep your operating system up to date and consult your system manager.
    Time to time look into the logs of WS and of the web server.
    
    \item
    A Fortran 95 compiler.
    It is only necessary, if the binaries provided fail to work, or you want to fine-tune the software to your system.
    The Fortran programs don't exploit any special features of a particular "Fortran dialect", they are written in a portable way. Enabling optimalization and paralellization speeds up the program, but a too agressive usage of these features, however, can result false results or error messages, hence you should always compare the results with that obtained without optimization and paralellization.
    
    \item
    A PERL installation~\cite{PERL}
    
    \item
    The memory-, disc-, and CPU requirements depend on the intended use of the software.
    For a minimum installation almost any not too old server computer can do, with a couple hundred MB of free RAM and disk space, but when we plan a classroom use (several concurrent sessions), and/or more complicated calculations, then it is better to have at least 4 GB free disk space, 1 GB RAM, and a multi core CPU.
    
\end{enumerate}

\subsection{Required libraries and programs\label{sec:required}}

\begin{itemize}

    \item
    {\tt AEDR\_v\_1\_0}.
    It is the program~\cite{AEDR} for the solution of the stationary Schr{\"o}dinger equation
    ({\tt 2dsch}).

    \item
    {\tt FFTW} Fast Fourier Transform libarary\cite{FFTW}. It is called from the "sfnd" and "2dsch" Fortran programs.
    
    \item
    {\tt PGPLOT} Graphics Subroutine Library~\cite{PGPLOT}.
    It is used by the "ploti" Fortran program.
    
    \item
    {\tt NETPBM} Graphics Library~\cite{NETPBM}.
    This is a collection of command line image processing tools.
    It is used in the PERL components of WS.
    
    \item
    {\tt CGI} Library~\cite{CGIWikipedia} (cgi-lib.pl).
    It is used by the PERL programs to handle the CGI protocol.
    It is a single PERL file and it has to be present in the CGI directory (see Section \ref{sec:folders}).
    
    \item
    {\tt GIFSICLE}\cite{GIFSICLE}. It is a command line tool for creating GIF animations.
    
\end{itemize}

\subsection{Folder structure\label{sec:folders}}

Web-Schr{\"o}dinger uses the following directories:

\begin{itemize}

    \item
    "wsch" directory.
    This is the main folder of the program.
    It has to be created in such a place of the filesystem, where it is enabled for the web server to run CGI scripts.
    For example it can be a subfolder in the main CGI directory.
    Look for its location in the help of the web server, or ask the system manager.
    In Linux machines it is ofter "/usr/lib/cgi-bin".
    Most of the components of the program has to be present in the "wsch" directory.
    This includes the compiled Fortran programs,
    the CGI scrips (*.cgi),
    the PERL scripts (*.pl),
    the JavaScript programs (*.js), and a few auxiliary files.
    These are all static files, but a few read-write files, like the logs and the counters are also in this directory.
    
    \item
    "wsch/iro" directory.
    This is the workspace of WS, a subfolder of the "wsch" directory..
    Many intermediate files are created here during the operation of the program.
    This directory is purged time to time by the program, in order to prevent overgrowing, but in case of a lot of parallel sessions and/or complicated calculations its size can grow to several hundred MBs, or more.
    
    \item
    "wsch/examples" directory, a subfolder of the "wsch" directory.
    This contains the example project files, read by the "Load example" menu item.

    \item
    A directory, where the user manual is installed.
    This includes a HTML file and several image- and animation files.
    Being static files, these can be installed anywhere on the server, where the web server can reach them.
    
\end{itemize}

\subsection{Program files\label{sec:programs}}

{\em Calculation and data conversion programs} (written in Fortran).
These have to be present in the "wsch" directory (cf. Section \ref{sec:folders}) as executable binary files, with the names listed in {\em Tab.~\ref{tbl:fortran}}.
Each program have a few source files, which we provided in separate directories.
All of the compiled executable programs need general execute permission.

{\em binary programs}.
We provide the compiled Fortran programs as part of the installation material, in the "wsch" directory.
They are provided on "as is" basis, there is no guarantee, that they will work on a particular computer configuration.
In an ideal case these binaries may work on a Linux machine.
If not, new binaries should be complied using the source material.

\begin{table}
\caption{
Calculation and data conversion programs
}\label{tbl:fortran}
\begin{tabular}{ll}
\hline
Program name & Description\\
\hline
pot & Calculating of the $V(x,y)$ potential\\
sfnd & Preparing the initial wave packet, time evolution, calculating the observables\\
2dsch & Solution of the stationary Schr{\"o}dinger equation\\
orbs & Conversion routine for the "2dsch" results\\
ploti & Plotting of the $I(t)$ functions\\
\hline
\end{tabular}
\end{table}

{\em User interface programs} (written in PERL).
These "*.pl" files have to be copied to the "wsch" directory.
The function of the main PERL programs are given in {\em Tab.~\ref{tbl:perl}}.
The other PERL programs (not listed) are auxiliary routines.
All of them must have general "read" permission.

\begin{table}
\caption{
User interface programs
}\label{tbl:perl}
\begin{tabular}{ll}
\hline
Program name & Description\\
\hline
meni\_rut & Menu system\\
fili\_rut & "File" menu\\
chai\_rut & "Edit" menu\\
devi\_rut & Time development\\
orbi\_rut & Stationary states\\
resi\_rut & "Results" menu\\
lock      & File locking mechanism\\
\hline
\end{tabular}
\end{table}

{\em CGI programs}.
These "*.cgi" files have to be copied to the "wsch" directory.
These small programs provide an interface layer between the web server software and the PERL programs.
All of them must have general "execute" permission.

{\em Java scripts}.
These "*.js" files have to be copied to the "wsch" directory.
These small routines run on the client computer and check user input before transferring it to the server.
All of them must have general "read" permission.

{\em HTML scripts}.
These "*.html" files have to be copied to the "wsch" directory.
These are templates for the dynamic HTML pages created by the program.
All of them must have general "read" permission.

{\em Miscalleneous files}.
"nextjobid.dat" and "sch\_counter.dat" are two counters, with initial content of one.
The program increments these counters.
These two files must have general "write" permission.

{\em iro2 directory}. This is the workspace of the program, but some files have to be present here permanently: "lut\_minus.raw" and "lut\_plus.raw" are color look-up tables; "macska.jpg" is the cat logo.
This directory must have general "write" permission.cd 

{\em examples directory}.
Project files ("*.prj") used in the "File/Load example" menu.

\subsection{Installation procedure\label{sec:InstProc}}

\begin{enumerate}

    \item
    Check the system requirements (cf. Sec. \ref{sec:SystemRequirements})
    
    \item
    Download the program from here:
    
    {\tt http://www.nanotechnology.hu}
    
    {\tt /online/web-schroedinger/repo/web\_schrodinger\_3\_2.gz}
    
    \item
    Install the required libraries and auxiliary programs (cf. Sec. \ref{sec:required})
    
    \item
    UNZIP the installation GZIP file to somewhere in your file system.
    Three main folders will be created: "wsch", "html", and "source".
    
    \item
    Determine whether the provided binaries (cf. {\em Tab.~\ref{tbl:fortran}}) can run in your system
    
    \begin{itemize}
    
        \item
        The simplest way to check the binaries is to try and start them.
        This can be done by going into the "wsch" directory and issuing the "./NAME" command from a command prompt, where "NAME" is one of the names given in {\em Tab.~\ref{tbl:fortran}}.
        If the program replies with a Fortran message, asking for some input, then it probably runs OK.
        
        \item
        If the binaries won't run, then use the Fortran source programs provided in the subdirectories
        of the folder "src" and compile them. The "2dsch" directory does not contain source files,
	    but "diff" files, because we can not provide the source files for the AEDR program for copyright reasons.
	    Those Fortran programs are however, freely available from the Computer Physics Communications program library~\cite{AEDR}.
	    Download, GUnzip, and UnTar the bundle. We need only the "2dsch" directory. Some of the source files have to be slightly modified, as given in the "diff" files of the "src/2dsch" directory.

        Copy the resulting binaries with names given in {\em Tab.~\ref{tbl:fortran}}
        into the folder "wsch", with "execute" permissions.
        
    \end{itemize}
    
    \item
    Copy the "wsch" directory, together with its subdirectories and keeping the permissions, into its final place
    (cf. Sec. \ref{sec:folders})
    
    \item
    Copy the contents of the "help" directory to its final place
    (cf. Sec. \ref{sec:folders})
    
    \item
    Edit the URLs in the files in the "wsch" and "html" directories as follows.
    In the installation material the URL of the calculation server is a dummy address, "calcserver.somewhere.domain".
    This should be changed into the valid host name of your calculation server in files
    
    The URL of the HELP is given as "htmlserver.somewhere.domain".
    That should be changd into the valid host name of your HTML server.

\end{enumerate}

\section{Summary}

We have presented in this paper a client-server program for interactive solution of the two-dimensional Schr{\"o}dinger equation.
It can be used in a research environment to quick study of simple 2D quantum mechanical models and in an education environment to teach and practice quantum mechanics.
The main advantage of the software is that, after installing its server component, it can be used from several client computers paralelly, without installing anything on the client computers.
The typical run time is in the seconds range, which makes it possible for interactive and classroom use.

The client-server approach used in WS can be successfully applied in creating similar interactive simulation programs for other physical concepts, like Maxwell equations, classical mechanics, etc, because the basic framework is the same for many physical simulations: first we specify some input quantites, then run a simulation, and finally display the results. 
Adapting WS for simulating some another phenomena can easily be accomplished by changing the calculation programs, the graphics, and the input forms. 

\section{Acknowledgements}

The work has been supported by the European H2020 Graphene Core2 project no. 785219, Graphene Flagship and by the Belgian FNRS.
Helpful discussions with Ph. Lambin, L. P. Bir{\'o}, and
P. Vancs{\'o} are gratefully acknowledged.

\appendix

\section{Example library\label{sec:examples}}

The example project files are available from the "Files/Load example" menu.
{\em Tab.~\ref{tbl:ExamplesTimedev}} lists the examples provided for the time development calculation, {\em Tab.~\ref{tbl:ExamplesStationary}} list the examples provided for the time independent calculation.
Of course, after loading an example created for the time dependent calculation it is perfectly possible to launch a time independent calculation instead (and vice versa), but the examples for the time development calculation were developed to illustrate important phenomena in WPD (e.g. tunneling of a WP) and the examples for the time independent calculation were developed to illustrate interesting bound states (e.g. a diatomic molecule).

\begin{table}
\caption{
Examples provided for the time development calculation
}\label{tbl:ExamplesTimedev}
\begin{tabular}{ll}
\hline
Example file name & Description\\
\hline
band\_1d\_allowed & Wave packet of allowed energy passing through a periodic potential\\
band\_1d\_forbidden & Wave packet of forbidden energy reflected from a periodic potential\\
christmas & A quantum Christmas tree\\
gravity & Quantum analogue of a projectile motion\\
hardcore & Scattering of a wave packet on a circular hardcore potential\\
quantum\_revival & Demonstration of the quantum revival phenomenon\\
stm\_on\_nanotube & Simulation of Scanning Tunneling Microscope imaging of a carbon nanotube\\
tunneling\_oblique & Tunneling of a wave packet on an oblique potential wall\\
tunneling\_perpendicular & Tunneling of a wave packet on a perpendicular potential wall\\
\hline
\end{tabular}
\end{table}

\begin{table}
\caption{
Examples provided for the stationary calculation
}\label{tbl:ExamplesStationary}
\begin{tabular}{ll}
\hline
Example file name & Description\\
\hline
box & Eigenstates for a rectangular hard wall potential box\\
circle & Eigenstates for a circular hard wall potential box\\
harmosc\_2d & Harmonical oscillator in two dimensions\\
molecule & Model of a diatomic molecule\\
step & Eigenstates for a step potential\\
\hline
\end{tabular}
\end{table}

\section{Example I -- quantum tunneling\label{sec:TunnelingExample}}

In this section we present the example file {\em tunneling\_perpendicular} in detail.
This example illustrates one of the most basic phenomena of the quantum mechanics, tunneling, from the point of view of WPD.
{\em Tab.~\ref{tbl:tunneling}} lists the parameters of this calculation.

\begin{table}
\caption{
The example {\em tunneling\_perpendicular}
}\label{tbl:tunneling}
\begin{tabular}{lll}
\hline
Parameter name & parameter value & description\\
\hline
$N_x$, $N_y$ & 256 & Number of mesh points in $x$ and $y$ direction\\
$s_x$, $s_y$ & 76.8 \AA & Size of the calculation window in $x$ and $y$ direction ($\delta x,y$ = 0.3 \AA)\\
Type of potential & rectangle & 2D geometrical object where the potential is defined\\
$x_0$, $y_0$ & 0.0 \AA & Center of the rectangle\\
{\em width}, {\em height} & 76.8 \AA, 2.7 \AA & $x$ and $y$ sizes of the rectange\\
$V_0$ & 7.0 eV & Potential value inside the rectange\\
$x_i$, $y_i$ & 0.0 \AA, 18.0 \AA & Initial position of the wave packet\\
$a_x$, $a_y$ & 5.0 \AA & Width of the wave packet\\
$E_{kin}$ & 5.0 eV & Initial kinetic energy\\
{\em Angle} & -90.0$^o$ & Direction of the initial momentum ($-y$)\\
$y_{det}^{(1)}$ & 3.0 \AA & Position of the first detector line\\
$y_{det}^{(2)}$ & -3.9 \AA & Position of the second detector line\\
\hline
\end{tabular}
\end{table}

\begin{figure}
	\centering
		\includegraphics[width=6 cm]{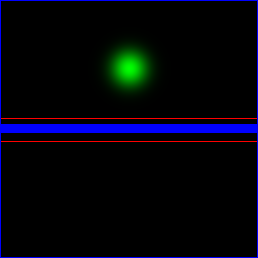}
	\caption{\label{fig:tunnelingsetup}
	The calculation setup of the {\em tunneling\_perpendicular} example.
	The middle blue bar is the tunneling potential.
	The upper green spot is the probability density of the intitial wave packet.
	The two red horizontal lines show the positions of the detectors.
	Size of the calculation window is $76.8$ \AA.
    }
\end{figure}

After loading the {\em tunneling\_perpendicular} project file from the "Files/Load example" menu item the user can make changes using the "Edit" menu, or go straight for the calculation.
{\em Fig.~\ref{fig:tunnelingsetup}} shows the setup of this calculation as a composite image of the potential, the initial WP, and the detector lines, as displayed by the "Edit/Initial state" menu item.

Upon clicking the "Calculation/Time development" menu item the program calculates snapshots of the WP probabilty density (cf. {\em Fig.~\ref{fig:tunnelingsnaps}}).

\begin{figure}
	\centering
		\includegraphics[width=12 cm]{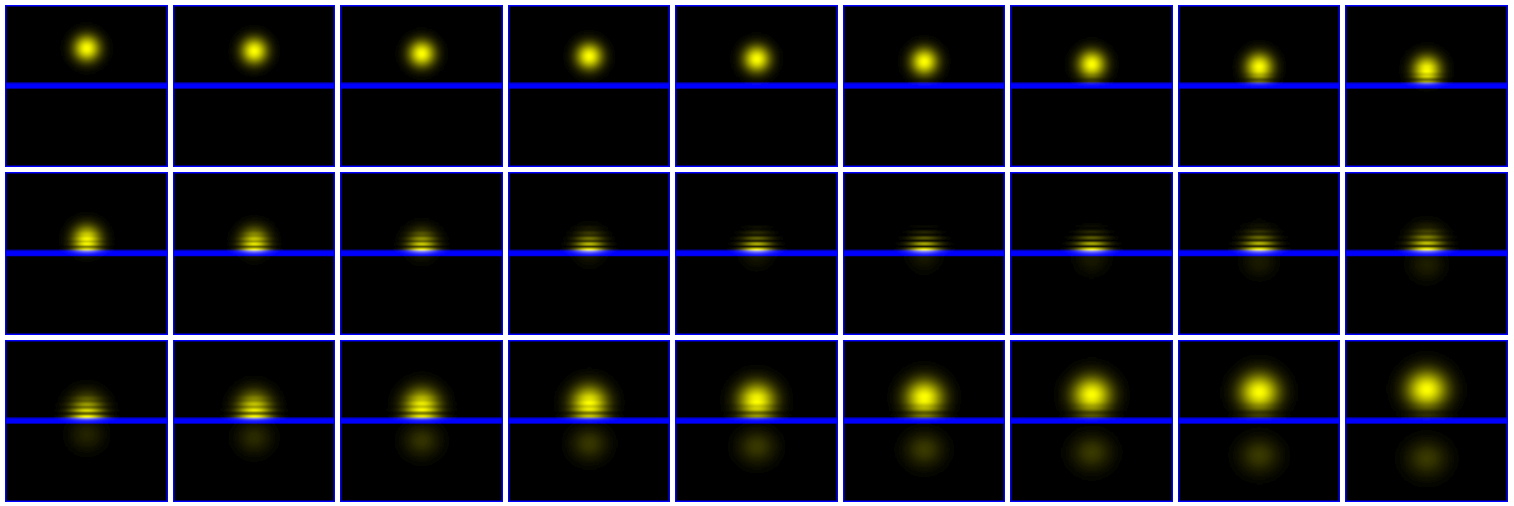}
	\caption{\label{fig:tunnelingsnaps}
	Snapshots of the time development of the wave packet probability density when running the {\em tunneling\_perpendicular} example.
	The display time step is 96.7 as (attosecond), the duration of the simulation is 2.4 fs (femtosecond).
	Note that the initial wave packet splits into a reflected and a transmitted part.
	Each of the snapshots are individually normalized.
	Size of the calculation window is $76.8$ \AA.
    }
\end{figure}

\section{Example II -- STM imaging of a carbon nanotube\label{sec:STMExample}}

In this section we present the example file {\em stm\_on\_nanotube}.
This example goes beyond the basic phenomena of quantum mechanincs, it is an application of WS to answer real scientific questions.
This calculation was published in detail in~\cite{Mark1998PRB} in 2D and later in~\cite{Mark2004PRB} in 3D.
The example also goes beyond the 1D quantum mechanical phenomena, the resonant states on the nanotube are 2D states~\cite{Tapaszto2006Apparent}.
As we explained in Section \ref{sec:InitialState}
these resonant states are caused by the presence of the two tunneling gaps.
After loading the example and
clicking the "Calculation/Time development" menu item the program calculates snapshots of the WP probability density (cf. {\em Fig.~\ref{fig:tunnelingsnaps}}), together with the $I(t)$ probability currents and $T$ transmission and reflection probabilities.
Computation of the transmission probability makes it possible to calculate the tunneling current in the STM.

\begin{figure}
	\centering
		\includegraphics[width=12 cm]{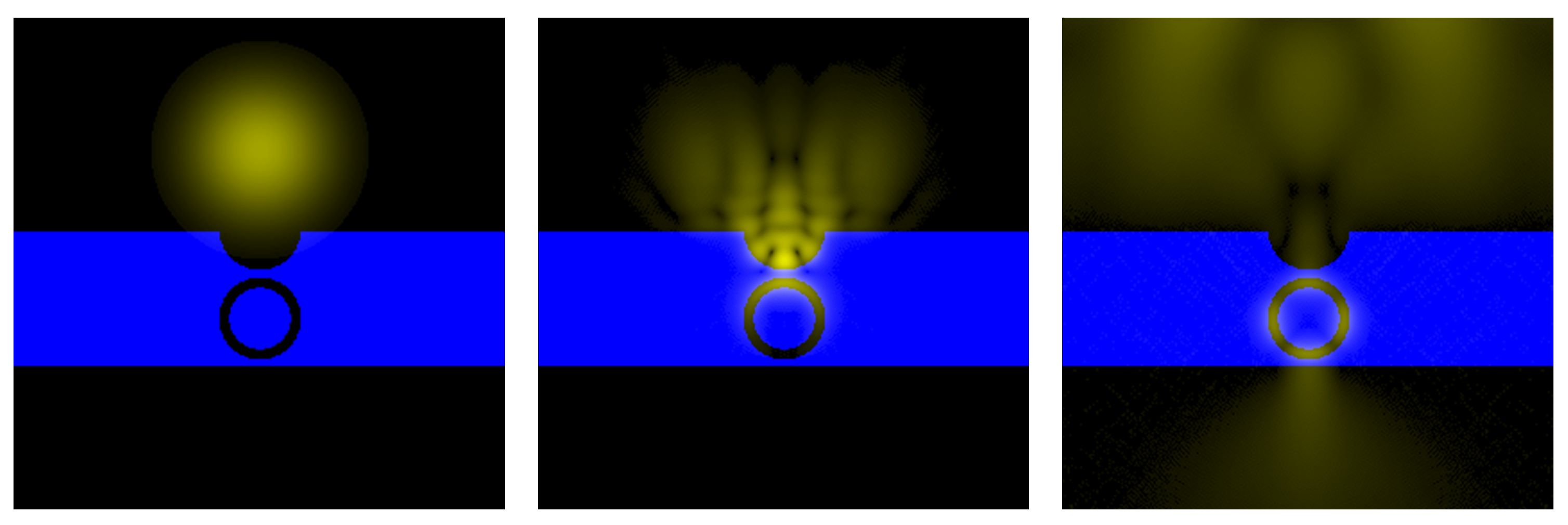}
	\caption{\label{fig:tunnelingsnaps}
	Snapshots of the time development of the wave packet probability density when running the {\em stm\_on\_nanotube} example
	for $t=0$ fs, $t=1.7$ fs, and $t=4.4$ fs.
	The upper semicircle is the cross section of the STM tip, the middle ring is the cross section of the nanotube.
	The wave packet starts from the bulk of the STM tip. then tunnels through the tip-nanotube and nanotube-support tunnel gaps.
	Note the interferences in the tip and nanotube.
	The middle blue bar is the
	tunneling potential barrier, blue is $V=0$ eV and black is $V=-9.81$ eV.
	Size of the calculation window is $12.8$ nm.
    }
\end{figure}

\bibliographystyle{elsarticle-num}

\bibliography{ws_2019_arxiv}


\end{document}